\documentclass[notitlepage,pre,amsfonts,floatfix,superscriptaddress,onecolumn,longbibliography]{revtex4-1}
\usepackage{mathbbol}
\usepackage{graphicx,color}
\usepackage{amsmath,amssymb,bm,amsthm}
\usepackage{mathtools}
\usepackage{simplewick}
\usepackage{braket}

\usepackage{mathrsfs}

\def\be{\begin{equation}}
\def\ee{\end{equation}}
\def\ba{\begin{eqnarray}}
\def\ea{\end{eqnarray}}

\begin{document}
\title{Dynamical Detection of Level Repulsion in the One-Particle Aubry-Andr\'e Model}

\author{E. Jonathan Torres-Herrera}
\address{Instituto de F\'isica, Benem\'erita Universidad Aut\'onoma de Puebla,
Apt. Postal J-48, Puebla, 72570, Mexico}
\author{Lea F. Santos}
\address{Department of Physics, Yeshiva University, New York City, NY, 10016, USA}

%\date{\today}

\begin{abstract}
The analysis of level statistics provides a primary method to detect signatures of chaos in the quantum domain. However, for experiments with ion traps and cold atoms, the energy levels are not as easily  accessible as the dynamics. In this work, we discuss how properties of the spectrum that are usually associated with chaos can be directly detected from the evolution of the number operator in the one-dimensional, noninteracting Aubry-Andr\'e model. Both the quantity and the model are studied in experiments with cold atoms. We consider a single-particle and system sizes experimentally reachable. By varying the disorder strength within values below the critical point of the model, level statistics similar to those found in random matrix theory are obtained. Dynamically, these properties of the spectrum are manifested in the form of a dip below the equilibration point of the number operator. This feature emerges at times that are experimentally accessible. This work is a contribution to a special issue dedicated to Shmuel Fishman.
\end{abstract}

\maketitle

\section{Introduction}

There has been a surprising revival of interest in quantum chaos, especially from a dynamical perspective, with the exponential growth of out-of-time ordered correlators (OTOC) taken as a main indication of  chaotic behavior~\cite{KitaevTalk,Maldacena2016PRD,Maldacena2016JHEP,Roberts2015PRL,Fan2017,Luitz2017b,Borgonovi2019,YanARXIV,GarciaARXIV}. The more traditional approach to quantum chaos, however, focuses on the properties of the spectrum and uses level statistics as in random matrix theory (RMT) as its main signature~\cite{Bohigas1986,Guhr1998,HaakeBook,StockmannBook}. There are several examples of cases where a correspondence between the exponential growth of the OTOC and level repulsion as in RMT has been found~\cite{Rozenbaum2017,Jalabert2018,RozenbaumARXIV,Chavez2019,Lewis-Swan2019}, but~exceptions also exist~\cite{Pappalardi2018,Hummel2019,CameoARXIV}. In the present work, we propose a way to directly detect the effects of level repulsion in the evolution of a quantum system. The quantity and model that we consider, namely, the number operator and the Aubry-Andr\'e model, are accessible to experiments with cold atoms~\cite{Schreiber2015}.

The Aubry-Andr\'e model has quasiperiodic disorder~\cite{Harper1955,Aubry1980,Sokoloff1985,CastroARXIV,Roy2019}, so is contrary to the Anderson model where the disorder is random, in one-dimension (1D), and for a single particle, it can present both localized and delocalized regimes. All states in the Aubry-Andr\'e model become localized only above a critical disorder strength, while in the one-particle 1D infinite Anderson model, all states are localized for any disorder strength~\cite{Anderson1958,Lee1985,Kramer1993,AndersonPhysicsToday}. Despite this difference, when the systems are {\em finite} and have small disorder strengths, they present similar level spacing distributions; namely, they show distributions as in RMT, the so-called Wigner--Dyson distributions~\cite{Sorathia2012,Torres2019}. This is a finite-size effect, not a signature of chaos. Wigner--Dyson distributions in these non-chaotic 1D models emerge when the localization length is larger than the system size. However, these models can still be used as a way to demonstrate how the properties of the spectrum get manifested in the dynamics of realistic quantum systems. Here, we show how the level repulsion present in the finite one-particle 1D Aubry-Andr\'e model affects its dynamics.

In studies of many-body quantum systems, it has been shown that the survival probability, that is, the probability to find the system in its initial state later in time, decays below its saturation value in systems that present level repulsion~\cite{Leviandier1986,Guhr1990,Wilkie1991,Alhassid1992,Gorin2002,Torres2017,Torres2017Philo,Torres2018,Cotler2017,Numasawa2019}. This dip below saturation is commonly known as correlation hole~\cite{Leviandier1986,Guhr1990,Wilkie1991,Alhassid1992,Gorin2002,Torres2017,Torres2017Philo,Torres2018}. In many-body quantum systems, the time for its appearance grows exponentially with system size~\cite{Schiulaz2019}, which makes its experimental observation very challenging even for relatively small systems. To circumvent this issue, one could employ systems with few-degrees of freedom~\cite{Torres2019,Lerma2019}. However, two other problems remain: the correlation hole in systems with many particles emerges at extremely low values of the survival probability, and this quantity is non-local in real space, while experiments usually deal with local quantities (exceptions include~\cite{SinghARXIV}). 

To solve these problems, we consider the one-particle 1D Aubry-Andr\'e model and study the evolution of the number operator. This is a local quantity routinely measured in experiments with cold atoms. In the presence of level repulsion, a correlation hole develops at times that grow just sublinearly with the system size. In addition, for systems that are not too large, the minimum point of the hole occurs at values that are not very small, and therefore, do not require extraordinary precision for detection. All these factors should make the experimental observation of the correlation hole viable in this model. 

Before proceeding with the presentation of our results, we note that this work is a contribution to a special issue  dedicated to  Shmuel Fishman. As such, we find it pertinent to mention that Griniasty and Fishman studied a generalization of the Aubry-Andr\'e model in ~\cite{Griniasty1988}. We expect our results to be valid in this broader picture also.

%%%%%%%%%%%%%%%%%%%
\section{Finite One-Particle One-Dimension Aubry-Andr\'e Model}

We study the one-particle 1D Aubry-Andr\'e model with open boundaries described by the following Hamiltonian,
\begin{equation}
H = \sum_{j=1}^{L} h \cos[(\sqrt{5} -1) \pi j + \phi] c_j^{\dagger} c_j - J \sum_{j=1}^{L-1} (c_j^{\dagger} c_{j+1} + c_{j+1}^{\dagger} c_{j} ).
\label{Eq:AA}
\end{equation}
Above, $c_j^{\dagger}$ $(c_j)$ is the creation (annihilation) operator on site $j$. The first term defines the quasiperiodic onsite energies with disorder strength $h$; $\phi$ is a phase offset chosen randomly from a uniform distribution $[0,2\pi]$; the second term is responsible for hopping the particle along the chain (we~choose $J=1$),  and $L$ is the number of sites. 

The basis vectors $|\varphi_j \rangle$ that we use to write the Hamiltonian matrix correspond to states that have the particle placed on a single site $j$, such as $|1000\ldots\rangle$. The eigenvalues of the matrix are denoted by $E_{\alpha}$ and the corresponding eigenstates are  $|\psi_{\alpha} \rangle = \sum_j C^{(j)}_{\alpha} |\varphi_j \rangle$, where  $C^{(j)}_{\alpha} = \langle \varphi_j |\psi_{\alpha} \rangle   = \left(C^{(j)}_{\alpha}\right)^* =   \langle \psi_{\alpha}  | \varphi_j \rangle $.

\subsection{Level Statistics}

To study the degree of short-range correlations between the eigenvalues, we consider the level spacing distribution $P(s)$, which requires unfolding the spectrum~\cite{Guhr1998,MehtaBook}, and  the ratio $\tilde{r}_\alpha$ between neighboring levels~\cite{Oganesyan2007,Atas2012}, which does not require unfolding the spectrum. To detect long-range correlations, we look at the level number variance~\cite{Guhr1998,MehtaBook}, which also requires unfolding the spectrum.

The unfolding  procedure consists of locally rescaling the energies. The number of levels with energy less than or equal to a certain value $E$ is given by the staircase function $N(E) = \sum_n \Theta (E-E_n)$, where $\Theta$ is the unit step function. $N(E)$ has a smooth part $N_{sm}(E)$, which is the cumulative mean level density, and a fluctuating part $N_{fl}(E)$. By unfolding the spectrum, one maps the energies $\{ E_1, E_2, \ldots \}$ onto $\{ \epsilon_1, \epsilon_2, \ldots \}$, where $\epsilon_n=N_{sm}(E_n)$, so that the mean level density of the new energy sequence becomes one.
Statistics that measure long-range correlations are more sensitive to the unfolding procedure than short-range correlations~\cite{Gomez2002}. In this paper, we discard 20\% of the energies from the edges of the spectrum, and obtain
$N_{sm}(E)$ by fitting the staircase function with a polynomial of degree~7.

\subsubsection{Short-Range Correlations}

In the spectra of full random matrices, neighboring levels repel each other and $P(s)$ follows the Wigner--Dyson distribution. The exact form of the distribution depends on the symmetries of the Hamiltonian.
\begin{equation}
P_{\text{WD}}(s)=a_{\beta} s^{\beta} \exp(- b_{\beta} s^2)
\end{equation}
has $\beta=1$ for the Gaussian orthogonal ensemble (GOE), where the full random matrices are real and symmetric; $\beta=2$ for the Gaussian unitary ensemble (GUE), where the full random matrices are Hermitian; and $\beta=4$ for the Gaussian symplectic ensemble (GSE), where the full random matrices are written in terms of quaternions. The values of the constants for  $a_{\beta}$ and $b_{\beta}$ are found, for example, in~\cite{MehtaBook}. The degree of correlation between the eigenvalues increases from GOE to GUE to GSE.

In contrast with the spectra of RMT, one may find systems with uncorrelated eigenvalues, where the level spacing distribution is Poissonian and systems with eigenvalues that are more correlated than in random matrices and nearly equidistant, as in the "picket-fence"-kind of spectra~\cite{Berry1977,Pandey1991} and the Shnirelman's peak~\cite{Chirikov1995}.

The ratio $\tilde{r}_\alpha$ between neighboring levels is defined as~\cite{Oganesyan2007,Atas2012}
\begin{equation}
\tilde{r}_\alpha = \min \left( r_\alpha, \frac{1}{r_\alpha}\right), \quad \text{where} \quad r_\alpha=\frac{s_\alpha}{s_{\alpha-1}},
\end{equation}
and $s_\alpha = E_{\alpha+1} - E_\alpha$ is the spacing between neighboring levels. The average value $\langle \tilde{r}\rangle $ over all eigenvalues varies as follows: $\langle \tilde{r}\rangle\approx 0.39 $ for the Poissonian distribution, $\langle \tilde{r}\rangle\approx 0.54 $ for the GOE, $\langle \tilde{r}\rangle\approx 0.60 $ for the GUE, $\langle \tilde{r}\rangle\approx 0.68 $ for the GSE, and $\langle \tilde{r}\rangle\approx 1 $ for picket-fence-like spectra.

\begin{figure}[h]
\centering
\includegraphics[width=1.2\columnwidth]{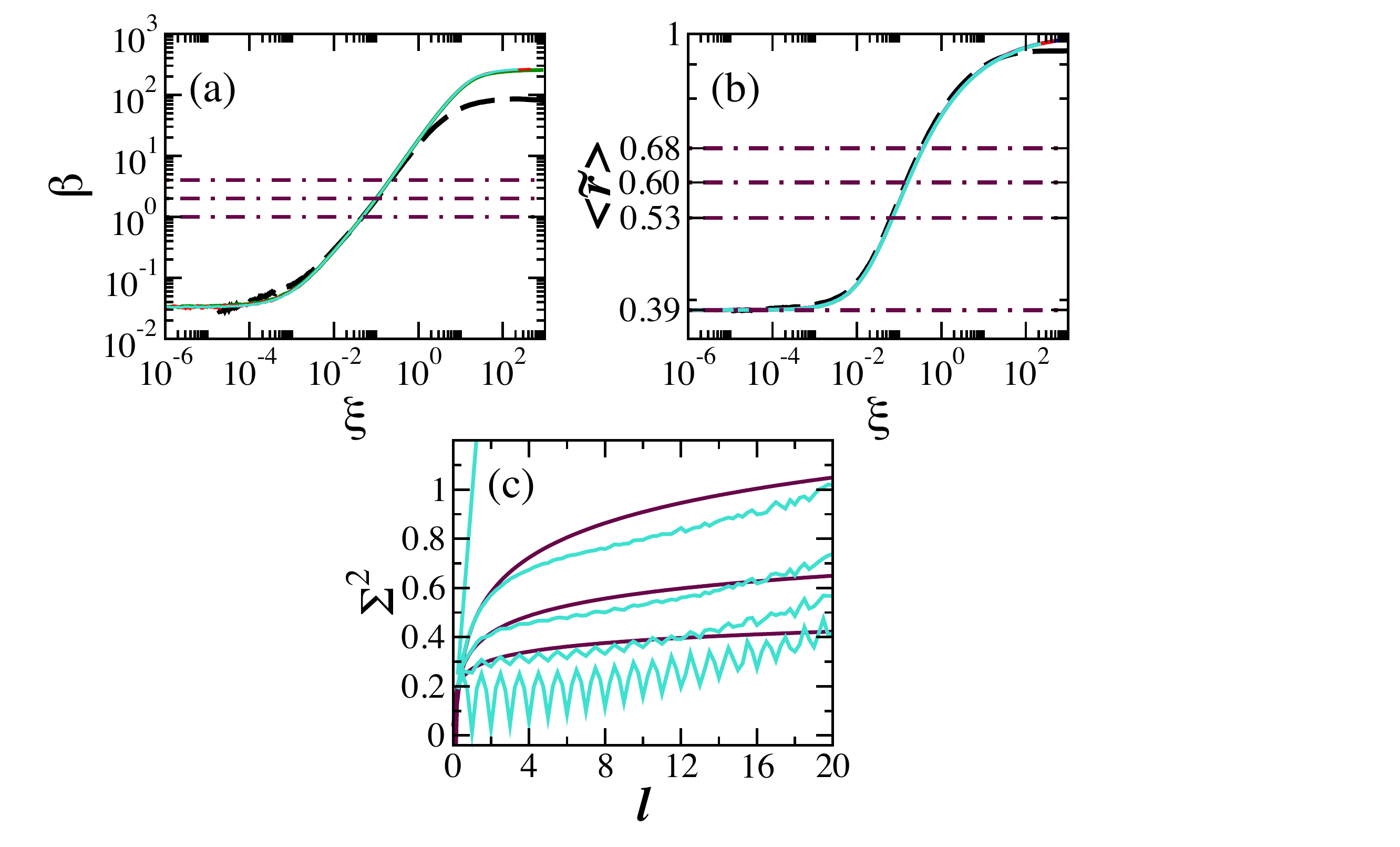}
\caption{Level repulsion parameter $\beta$ (\textbf{a}) and average ratio of spacings between consecutive levels $\langle \tilde{r}\rangle$ as a function of $\xi=1/(h^2 L)$, and level number variance (\textbf{c}). (\textbf{a},\textbf{b}) Four system sizes are considered, $L=100, 1000, 2000,$ and $4000$. The four curves overlap, except for the smallest one in panel (\textbf{a}). The~horizontal dot-dashed lines indicate the values for the Gaussian orthogonal ensemble (GOE), Gaussian unitary ensemble (GUE), and Gaussian symplectic ensemble (GSE) from top to bottom. (\textbf{c})~Numerical results for $L=4000$ (light color) and analytical curves for GOE, GUE, and GSE (dark color). The numerical curves from top to bottom have values of $\xi$ that, according to $\beta$, lead to the Poissonian distribution, GOE shape, GUE shape, GSE shape, and the picket-fence spectrum for $h=0$.  In all panels: averages over $10^3$ random realizations. }
\label{fig:beta}
\end{figure}

For the finite one-particle 1D Aubry-Andr\'e model,  the distribution is Poissonian  when $h$ is large. As the disorder strength decreases towards zero, where the eigenvalues become nearly equidistant, $P(s)$ passes through all forms mentioned above, from Poisson to GOE-like, from GOE-like to GUE-like, from GUE-like to GSE-like, and finally from GSE-like to the picket-fence case, with all the intermediate distributions between each specific case. This is shown in Figure~\ref{fig:beta}a,b. 

In Figure~\ref{fig:beta}a, we show the values of $\beta$ obtained with the expression~\cite{Izrailev1990}, %Layout, please fix callouts.
\begin{equation}
P_\beta (s) = A \left( \frac{\pi s}{2} \right)^\beta 
\exp \left[ -\frac{1}{4} \beta \left( \frac{\pi s}{2} \right)^2 - 
\left( B s -  \frac{\beta}{4} \pi s \right)  \right] ,
\label{Eq:beta}
\end{equation}
where $A$ and $B$ come from the normalization conditions
\begin{equation}
\int_{0}^{\infty} P_\beta (s)  ds = \int_{0}^{\infty} s P_\beta (s)  ds =1.
\end{equation}
The values of $\beta$ are shown as a function of the ratio $\xi=1/(h^2 L)$. This scaling factor collapses the curves for different system sizes on a single curve.
In Figure~\ref{fig:beta}b, we depict $\langle \tilde{r}\rangle$ as a function of $\xi$. While~both $\beta$ and $\langle \tilde{r}\rangle$ capture the crossovers from the Poissonian distribution up to the picket-fence spectrum as $\xi$ increases, it is evident that there is not an exact one-to-one correspondence between the two, but a more systematic comparison of the two quantities together with a careful unfolding procedure is worth doing. In this case, various different models should be taken into account, including true chaotic models.

It is important to emphasize that the different level spacing distributions obtained with the model are not linked with the symmetries of the Hamiltonian. The Hamiltonian matrix used here is real and symmetric for any value of $h\geq0$. The different forms of the distributions are rather a consequence of the changes in the level of correlations as one goes from uncorrelated eigenvalues for large disorder to nearly equidistant levels for the clean chain.

There are other theoretical studies where level statistics as in RMT were generated~\cite{Wu1993,Zhang1995,Relano2004}. Those~approaches are different from the one taken in the present work, where we do not build the matrix elements with the purpose of generating specific level statistics; instead, they emerge due to finite size effects.   

\subsubsection{Long-Range Correlations}

The analysis of long-range correlations can be done with  the level number variance; that is, the~variance $\Sigma^2 (\ell) $ of the unfolded levels in the interval $\ell$. For uncorrelated eigenvalues, $\Sigma^2 (\ell) $ grows linearly with $\ell$. In the case of full random matrices, we have for the  GOE~\cite{Guhr1998}, 
\begin{equation}
\Sigma^2_1 (\ell) = \frac{2}{\pi^2} \left( \ln(2 \pi \ell) + \gamma_e + 1 -\frac{\pi^2}{8} \right) ,
\end{equation}
for the GUE,
\begin{equation}
\Sigma^2_2 (\ell) = \frac{1}{\pi^2} \left( \ln(2 \pi \ell) + \gamma_e + 1 \right) ,
\end{equation}
and for the GSE,
\begin{equation}
\Sigma^2_4 (\ell) = \frac{1}{2\pi^2} \left( \ln(4 \pi \ell) + \gamma_e + 1 +\frac{\pi^2}{8} \right) ,
\end{equation}
where $\gamma_e = 0.5772\ldots $ is Euler's constant. For equidistant levels, as in the case of the harmonic oscillator, $ \Sigma^2 (\ell) =0$.

The plot of $\Sigma^2 (\ell) $ in Figure \ref{fig:beta}c~%newly added information, please confirm.
 makes it clear that the level of rigidity of the spectrum of the finite one-particle 1D Aubry-Andr\'e model is not equivalent to that for full random matrices. There is agreement for very small $\ell$, but then, for an interval of values of $\ell$, the correlations are stronger in the Aubry-Andr\'e model, until this behavior switches at large values of $\ell$ (compare the light and dark curves). As for the picket-fence spectrum for the clean chain (bottom light curve), we attribute the oscillations and the latter growth with $\ell$ to imperfections in the unfolding procedure and in the calculation of the level number variance, and to the fact that the eigenvalues are not exactly equidistant.

\section{Evolution of the Number Operator}

\begin{figure}[h]
\centering
\includegraphics[width=1\columnwidth]{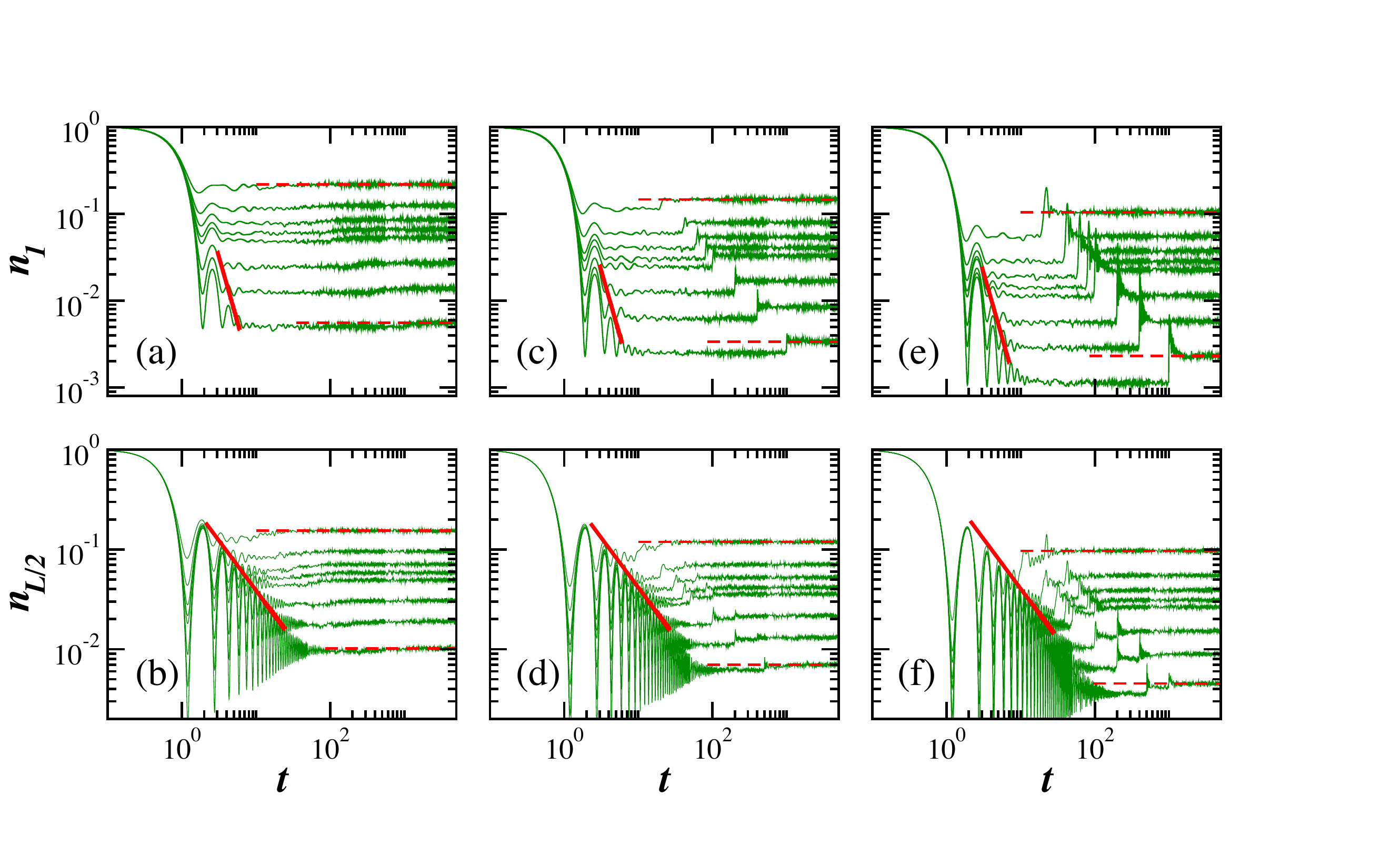}
\caption{Evolution of the number operator for a particle initially placed on Site 1 (\textbf{a},\textbf{c},\textbf{e}) and for one initially placed on Site $L/2$ (\textbf{b},\textbf{d},\textbf{f}) respectively, for spectra with GOE- (\textbf{a},\textbf{b}), GUE- (\textbf{c},\textbf{d}), and GSE-like (\textbf{e},\textbf{f}) level spacing distributions. On each panel, the size $L$ of the chain increases from top to bottom, $L=20,40,60,80,100,200,400,1000$.  The red solid lines represent the power-law decays: $1/t^3$ for (\textbf{a},\textbf{c},\textbf{e}) and $1/t$ for (\textbf{b},\textbf{d},\textbf{f}). The horizontal dashed lines mark the saturation values for the smallest and largest $L$'s. In all panels: averages over $10^3$ random realizations.
}
\label{fig:SP}
\end{figure}

Let us prepare the system in a state $|\Psi(0) \rangle =|\varphi_{j_0}\rangle$, where the particle is either on the first site of the chain, $j_0=1$, or on the middle one, $j_0=L/2$. We then evolve it under $H$ (\ref{Eq:AA}), $|\Psi(t) \rangle = e^{-i H t} |\Psi(0)\rangle$. The quantity used in the analysis of the dynamics is the number operator, 
\begin{equation}
n_{1,L/2} (t) = \langle \Psi(t) | c_{1,L/2}^{\dagger} c_{1,L/2} |\Psi(t) \rangle.
\end{equation}
The results for $n_1(t)$ and $n_{L/2}(t)$ are shown in Figure~\ref{fig:SP} on the top [(a), (c), (e)] and bottom (b), (d), (f)] panels, respectively. In Figure~\ref{fig:SP}a,b, the value of $\xi$ leads to the GOE-like level spacing distribution; in~Figure~\ref{fig:SP}c,d the distribution is GUE-like, and in Figure~\ref{fig:SP}e,f is GSE-like.

The main result of Figure~\ref{fig:SP} is the fact that for experimental sizes (few dozens of sites), the correlation hole emerges at times ($t<10^2$) and values of the number operator ($n_{1,L/2} (t)>10^{-2}$) that are experimentally reachable. The correlation hole is the dip below the saturation point of the dynamics. In all panels of Figure~\ref{fig:SP}, the saturation of the dynamics is marked with a red horizontal dashed line for the smallest and the largest system sizes. The correlation hole corresponds to the values of the numerical curves that are below this dashed line. The difference between saturation and minimum of the hole is most evident for the GSE-like spectrum in Figure~\ref{fig:SP}e.

One can write the number operator in terms of the energy eigenstates and eigenvalues as
\begin{equation}
n_{1,L/2} (t) = \left|\sum_{\alpha} |C_{\alpha}^{(1,L/2)} |^2 e^{-i E_{\alpha} t}  \right|^2 = 
 \left| \int_{E_{min}}^{E_{max}} \left(\sum_{\alpha} |C_{\alpha}^{(1,L/2)} |^2 \delta(E-E_{\alpha}) \right) e^{-i E t}  dE \right|^2 ,
 \label{Eq:n}
\end{equation}
where  $E_{min}$ is the lower bound of the spectrum and $E_{max}$ is the upper bound. In the equation above, the sum in parenthesis,
\begin{equation}
\rho_{1,L/2} = \sum_{\alpha} |C_{\alpha}^{(1,L/2)} |^2 \delta(E-E_{\alpha}),
\end{equation}
is the energy distribution of the initial state, often known as local density of states (LDOS) or strength function~\cite{Flambaum1994,Angom2004,Izrailev2006,Torres2014PRA}. The number operator in Equation~(\ref{Eq:n}) is the square of the Fourier transform of the LDOS. We denote the variance of the LDOS by 
\begin{equation}
\sigma_{1,L/2}^2 = \sum_{j\neq j_0} |\langle \varphi_j |H| \varphi_{j_0} \rangle|^2.
\end{equation} 

\begin{figure}[h]
\centering
\includegraphics[width=1\columnwidth]{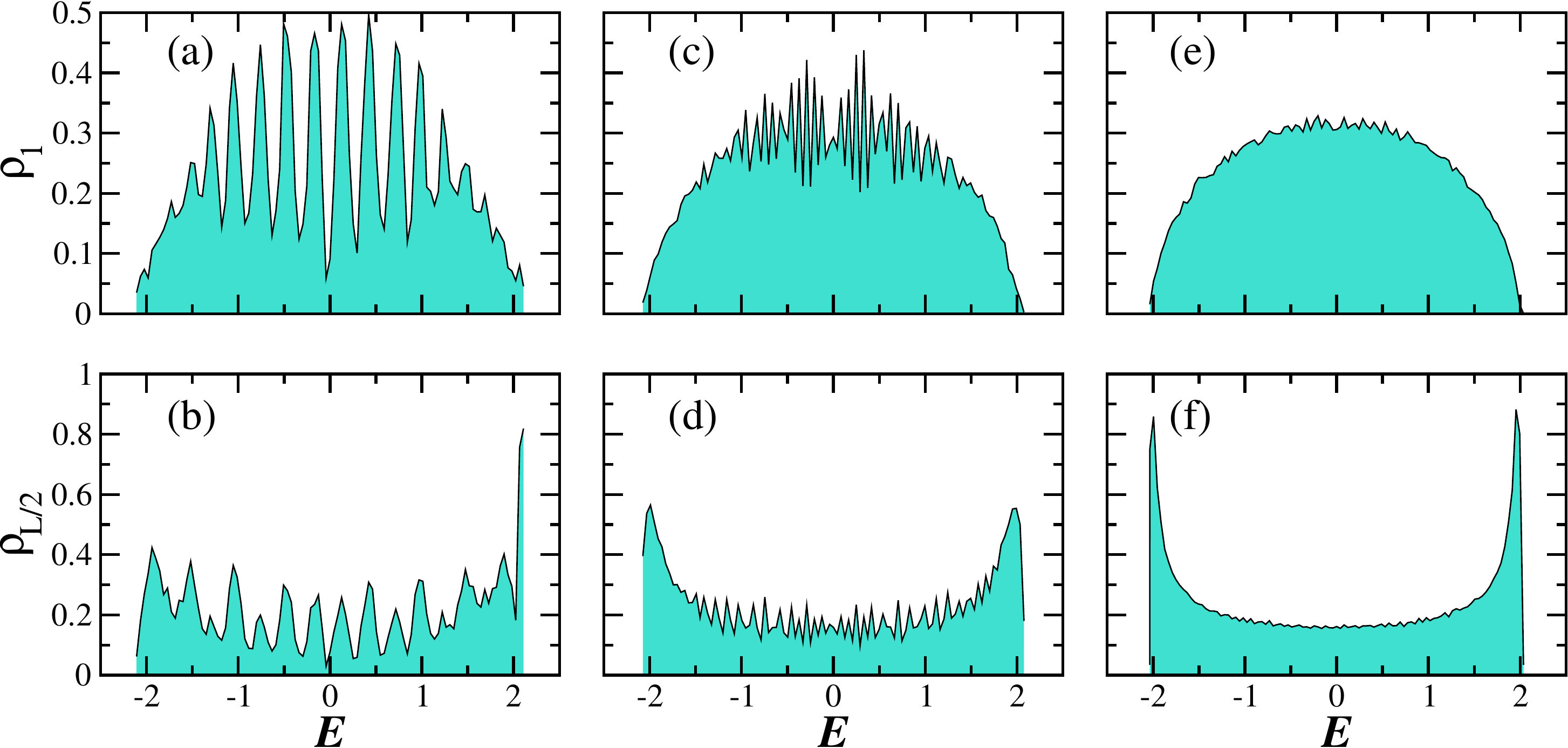}
\caption{Energy distribution of the initial state (LDOS) for a particle initially on Site 1 (\textbf{a},\textbf{c},\textbf{e}) and on Site $L/2$ (\textbf{b},\textbf{d},\textbf{f}) for spectra with  GSE-like level spacing distribution. The sizes of the chain increase from left to right: $L=20$ (\textbf{a},\textbf{b}); $L=80$ (\textbf{c},\textbf{d}), $L=400$ (\textbf{e},\textbf{f}). Average over $10^3$ random realizations.}
\label{fig:LDOS}
\end{figure}

The envelope of the LDOS for the Hamiltonian with GOE-, GUE-, and GSE-like level spacing distributions is analogous to the shape obtained for the clean model~\cite{Torres2019}. It is a semicircle when $j_0=1$ and it has a U-shape when $j_{L/2}=1$, as shown in Figure~\ref{fig:LDOS} for the GSE-like spectrum and three system sizes increasing from left to right, $L=20, 80, 400$. For small sizes, such as $L=20$, one sees approximately $L/2$ peaks in the middle of the LDOS. As $L$ increases, the curves become smoother.

The Fourier transform of the semicircle gives
\begin{equation}
n_1 (t) =\frac{[{\cal J}_1 (2  \sigma_1 t)]^2 }{ \sigma_1^2 t^2}, \hspace{1 cm } \mbox{where}
\hspace{1 cm } \sigma_1^2 = 1 ,
\end{equation}
and ${\cal J}_1$ is the Bessel function of the first kind.
For the U-shaped LDOS, we get
\begin{equation}
n_{L/2} (t) = [{\cal J}_0 (2 \sigma_{L/2} t)]^2, \hspace{1 cm } \mbox{where}
\hspace{1 cm } \sigma_{L/1}^2 = 2.
\end{equation}
The equations above imply that the initial decay of $n_1 (t<\sigma_1) \approx 1 - t^2$ is slower than for $n_{L/2} (t<\sigma_{L/2})  \approx 1 - 2 t^2$, which is noticeable by comparing the top and bottom panels of Figure~\ref{fig:SP} for $t< 1$. This is expected, since the particle on Site 1 can only hop to Site 2, while Site $L/2$ has two neighbors. 

For $t>\sigma_{1,L/2}$, the picture changes and the dynamics become faster for $n_1(t)$ than for $n_{L/2} (t)$. The quadratic decay is succeeded by a power-law decay that envelops the oscillations of the Bessel functions. This non-algebraic decay $\propto 1/t^{\gamma}$ is caused by the bounds in the spectrum~\cite{Tavora2016,Tavora2017}. The exponent is $\gamma=3$ for $n_1 (t)$ \cite{TorresKollmar2015} and $\gamma=1$ for $n_{L/2} (t)$ \cite{Santos2016}. 

The power-law decay is followed by a plateau that is below the saturation value,
\begin{equation}
\overline{n}_{1,L/2} = \sum_{\alpha} |C_{\alpha}^{(1,L/2)} |^4,
\end{equation}
of the number operator.
This saturation point is marked with dashed horizontal lines in Figure~\ref{fig:SP}. The plateau below this point corresponds to the correlation hole. It is related to the level number variance~\cite{Guhr1998,Alhassid1992}, which explains why it gets deeper as we move from the GOE- to the GUE- and to the GSE-like spectrum (compare Figure~\ref{fig:beta}c and Figure~\ref{fig:SP}).
The hole does not develop in integrable models where the level spacing distribution is  Poissonian and the eigenvalues are uncorrelated. But it does emerge in integrable models with a picket-fence spectrum. 

\begin{figure}[h]
\centering
\includegraphics[width=0.7\columnwidth]{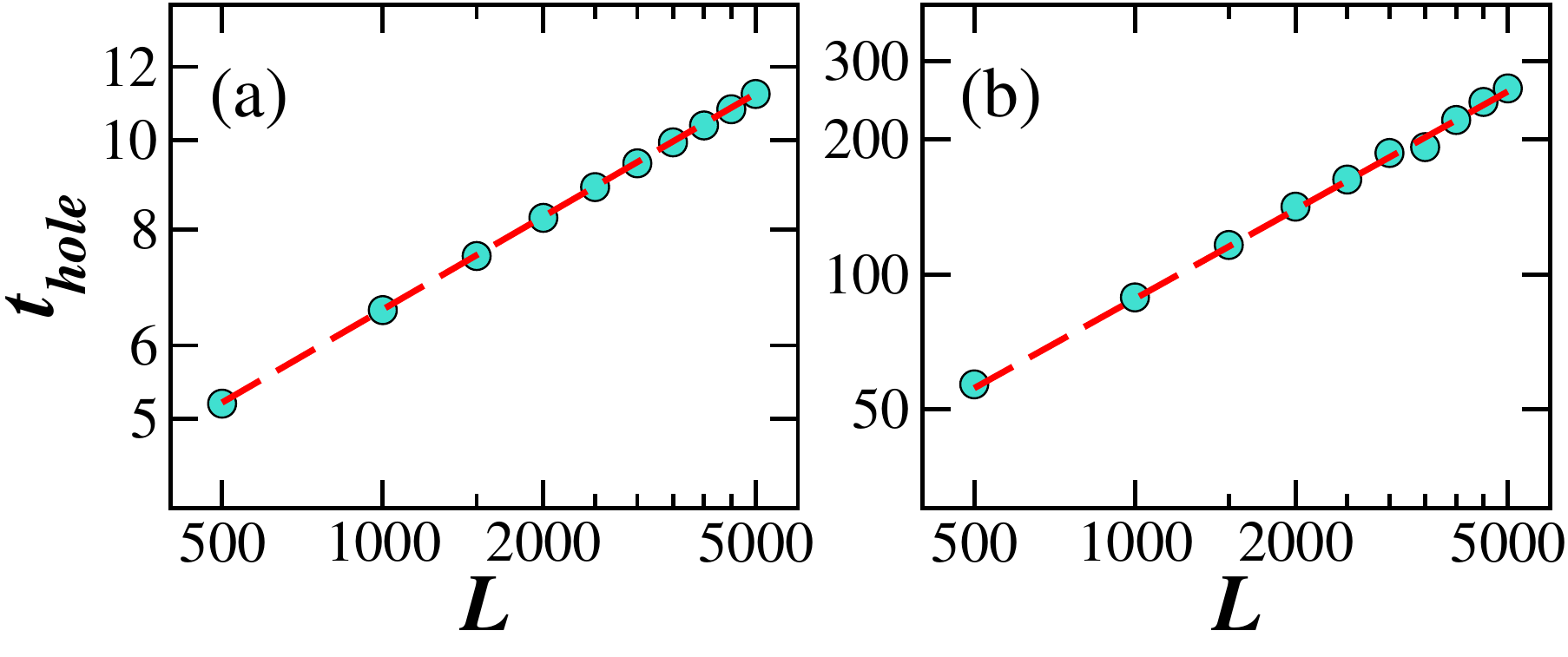}
\caption{Log--log plots for the time to reach the correlation hole for a particle initially placed on Site 1 (\textbf{a}) and a particle initially placed on Site $L/2$ (\textbf{b}) versus the system size for spectra with  GSE-like level spacing distributions. (\textbf{a}) $t_{\text{hole}} \propto L^{1/3}$ and (\textbf{b}) $t_{\text{hole}} \propto L^{2/3}$. Average over $10^3$ random realizations.}
\label{fig:thole}
\end{figure}

By checking where the curve of the power-law decay first crosses the plateau below $\overline{n}_{1,L/2}$, we~estimate numerically, the time $t_{\text{hole}}$ for the minimum of the correlation hole. As shown in Figure~\ref{fig:thole}, we find that $t_{\text{hole}} \propto L^{1/3}$ for $n_{1}(t)$ and $t_{\text{hole}} \propto L^{2/3}$ for $n_{L/2}(t)$. The first estimate can be derived from the fact that the power-law decay is $\propto 1/t^3$ and the minimum value of $n_{1}(t)$ at the plateau is $\propto 1/L$. The estimate for the $t_{\text{hole}}$ for $n_{L/2}(t)$ comes from the power-law decay $\propto 1/t$ and the minimum value of $n_{L/2}(t)$ at the plateau, which is $\propto 1/L^{2/3}$. Both times should be reachable by current experiments with cold atoms realized with few dozens of sites.

The correlation hole holds up to the revival of the dynamics, which first happens at $t_{rev} \sim L$ for $n_{1}(t)$ and at $t_{rev} \sim L/2$ for $n_{L/2} (t)$, as seen in Figure~\ref{fig:SP}. The revival is followed by another decay and a possible correlation hole, but at higher values. This behavior is better seen for the GSE-like spectrum in Figure~\ref{fig:SP}f, where the correlation is deep. The revival repeats itself at $t_{rev} \sim 2 L$ for $n_{1}(t)$ and at $t_{rev} \sim L$ for $n_{L/2} (t)$ with an yet larger value of the correlation hole. This second revival is better seen for larger $L$'s. We may expect subsequent revivals to become visible to even larger system sizes, although they should eventually become indistinguishable of the temporal fluctuations at the saturation point.

%%%%%%%%%%%%%%%%%%%
\section{Conclusions}

This work shows that the effects of level repulsion can be directly observed by studying the evolution of the number operator in the finite one-particle 1D Aubry-Andr\'e model. Level repulsion is manifested in the form of the so-called correlation hole. The number operator, the Aubry-Andr\'e model, the system sizes, and timescales studied here are accessible to experiments with cold atoms.

%%%%%%%%%%%%%%%%%%%%%%%%
\begin{acknowledgments}
E.J.T.-H. acknowledges funding from VIEP-BUAP
(grant numbers MEBJ-EXC19-G and LUAG-EXC19-G). L.F.S. was supported by the NSF, grant number~DMR-1936006. E.J.T.-H. is grateful to LNS-BUAP for allowing use
of their supercomputing facility.
\end{acknowledgments}
%%%%%%%%%%%%%%%%%%%%%%%%

%%%%%%%%%%%%%%%%%%%%%%%%
%

\end{document}